\documentclass[preprint,10pt]{elsarticle}
\usepackage{amssymb}
\usepackage[T2A]{fontenc}
\usepackage[utf8]{inputenc}
\usepackage{wrapfig}
\usepackage[]{graphicx,xcolor}
\usepackage{tabularx}
\usepackage{booktabs}
\usepackage{textcomp}
\usepackage{amsmath}
\usepackage{comment}
\usepackage[]{graphics}
\usepackage{amssymb}
\usepackage{amsmath}
\usepackage{color}
\usepackage{ifpdf}
\usepackage{lipsum}
\usepackage{siunitx}
\usepackage{braket}
\usepackage{lipsum}
\usepackage{soul}
\newcommand{\edit}[1]{{\color{black}{#1}}}
\newcommand{\editd}[1]{{\color{black}{#1}}}

\usepackage[margin=0.8in]{geometry}
\usepackage{float}
\usepackage{appendix}

\journal{Photonics and Nanostructures: Fundamentals and Applications}

\begin{document}

\begin{frontmatter}

\title{Resonant mode approximation of the scattering matrix of photonic crystal slabs near several Wood-Rayleigh anomalies}

\author[inst1,inst2]{D.\,A.\,Gromyko}
\author[inst1]{S.~A.~Dyakov}
\author[inst3]{V.\,A.\,Zinovyev}
\author[inst2,inst4]{S.~G.~Tikhodeev}
\author[inst1]{N.~A.~Gippius}

\affiliation[inst1]{organization={Skolkovo Institute of Science and Technology},
            addressline={Nobel Street 3}, 
            city={Moscow},
            postcode={143025}, 
            country={Russia}}
\affiliation[inst2]{organization={Faculty of Physics, Lomonosov Moscow State University},
            addressline={Leninskie Gory, d.1, str.2}, 
            city={Moscow},
            postcode={119991}, 
            country={Russia}}
\affiliation[inst3]{organization={Rzhanov Institute of Semiconductor Physics, SB RAS},
            addressline={prospekt Lavrent’eva 13}, 
            city={Novosibirsk},
            postcode={630090}, 
            country={Russia}}
\affiliation[inst4]{organization={A.~M.~Prokhorov General Physics Institute, RAS},
            addressline={Vavilova 38}, 
            city={Moscow},
            postcode={119991}, 
            country={Russia}}

\begin{abstract}The resonant mode approximation of the scattering matrix is considered for calculating the optical properties of multilayered periodic structures within the formalism of the Fourier-modal method for two diffraction thresholds in close proximity of the spectral-angular range of interest. The developed approximation opens up possibilities for the fast calculation of the scattering matrix of these structures when describing the integral characteristics of spectra and dispersion curves containing high-Q resonances, such as bound states in the continuum.\end{abstract}




\end{frontmatter}

\section{Introduction}
Multilayered periodic structures are of great interest in optics and photonics since they provide excellent opportunities for controlling electromagnetic waves with a wavelength comparable to the period. Theoretical and experimental studies of periodic structures over the past few decades have revealed many remarkable physical effects, many of which are of direct practical importance. The physical reason for a wide variety of new phenomena in photonic crystal layers is the diffraction of light on the periodic profile of the surface of such structures. If the photonic crystal layer is also a waveguide layer (that is, such that its average dielectric constant is greater than the dielectric constants of neighboring layers), then when a plane electromagnetic wave falls on it, it diffracts and excites quasiguided modes \cite{Tikhodeev2002b}. This phenomenon is fundamentally non-observable in the case of a planar multilayer structure without periodicity in the lateral directions.

An effective way to theoretically describe the whole variety of such effects in multilayered periodic structures is to use the Fourier-modal method in the form of a scattering matrix \cite{Tikhodeev2002b} also known as Rigorous coupled-wave analysis \cite{moharam1995formulation}. This method represents a mathematical formalism for solving Maxwell's equations in each of the layers of the structure (homogeneous or periodic) in the form of spatial Floquet-Fourier harmonics and connecting solutions in adjacent layers taking into account the boundary conditions. As a result of the iterative procedure proposed by Ko and Inkson \cite{ko88}, the scattering matrix of a structure $\mathbb{S}$ is calculated. By definition, the $\mathbb{S}$-matrix connects the input and output vectors of the amplitudes $|Out\rangle$ and $|In\rangle$ \edit{(see Fig. \ref{ScatteringMatFig})}:
\begin{figure}
    \centering
    \includegraphics[width=0.4\linewidth]{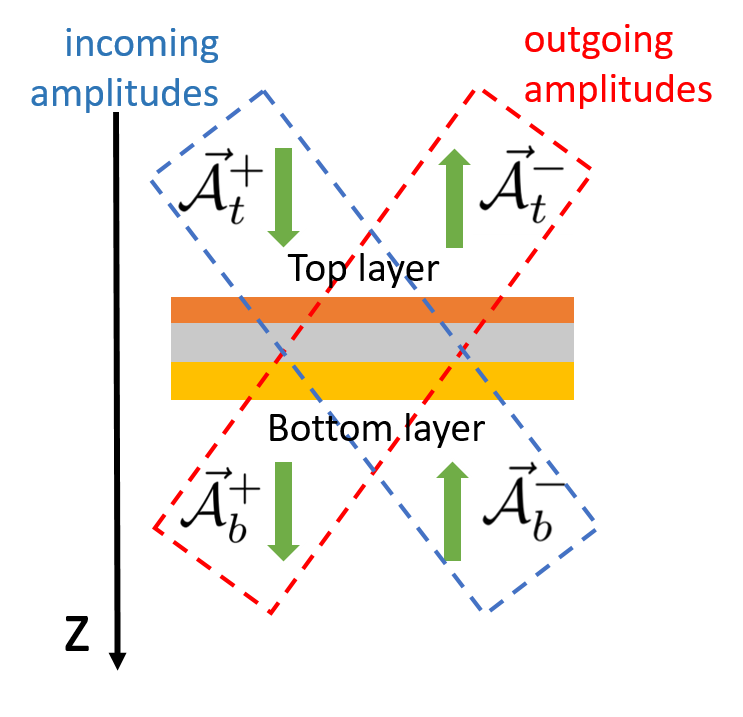}
    \caption{Schematic description of scattering matrix.}
    \label{ScatteringMatFig}
\end{figure}
\begin{equation}
    \mathbb{S} |In\rangle = |Out\rangle,
\end{equation}
{where $|In\rangle,\,|Out\rangle$ vectors consist of the amplitudes of the Fourier harmonics of electromagnetic fields taken at infinitesimally \edit{small} distance above the upper boundary and below the lower boundary of a model structure.}
This definition of the optical scattering matrix, given in Refs.\,\cite{whittaker1999scattering,Tikhodeev2002b}, is characterized by high computational stability and the ability to regularly carry out a formal procedure for its calculation for any arbitrarily complex structure at any given photon energies and wavevectors $k=(k_x,k_y,k_z)$. Despite this, when studying the optical properties of periodic structures by the Fourier-modal method, to obtain an accurate result, it is necessary to pay special attention to the convergence of the numerical scheme with respect to the number of harmonics used. Due to the slow convergence, early implementations of the Fourier modal method were good only for all-dielectric structures characterized by a weak dielectric contrast of periodic layers. However, additional techniques such as factorization rules \cite{li1997new, li2003fourier}, adaptive spatial resolution \cite{weiss2009matched, Weiss2011}, dipole approximation \cite{fradkin2019fourier}, and \edit{M}oiré adaptation \cite{salakhova2021fourier} improved the convergence and, thus, adapted the Fourier-modal method to calculating the scattering matrix of a much wider class of periodic structures, including metal-dielectric structures, structures with small metal particles \cite{fradkin2020nanoparticle}, structures with \edit{M}oiré superlattices \cite{salakhova2021fourier} etc.

The high degree of formalization of the Fourier-modal method allows it to be used as a tool for calculating not only such computationally simple characteristics as reflectance and transmittance, but also much more complex ones, for example, the Purcell factor of a radiating oscillating dipole in an inhomogeneous dielectric environment \cite{lobanov2012emission}, the power of heat transfer between periodic structures at near-field radiation heat transfer \cite{PhysRevB.85.180301,lussange2012radiative,PhysRevB.94.125431, PhysRevB.93.155403}, the Casimir force and torque arising when two periodic gratings are located close to each other \cite{lambrecht2008casimir,davids2010modal,PhysRevA.86.062502,antezza2020giant,graham2014casimir}. The reason is that our ability to construct the scattering matrix at given photon energy and wavevector appears to be enough for the calculation of those exotic quantities. The difficulty lies in the fact that the mathematical expressions for the Purcell factor, the heat transfer power, and the Casimir force contain integrals over the first Brillouin zone of some functions of the wavevector and photon energy. Numerical integration is a routine procedure; however, it can be complicated because the function to be integrated contains a certain number (sometimes quite significant) of high-Q resonance peaks. Such peaks can be a) symmetry protected bound states in the continuum, caused by a mismatch of the symmetry types of the structure eigenmode and the mode propagating in free space, b) bound states in the continuum, caused by destructive interference of interacting resonances in the strong coupling regime (bound states in the continuum of the Friedrich-Wintgen type), c) quasiguided modes of a weakly absorbing structure, lying under the vacuum and substrate light cones, etc. Since the position of the narrow resonance lines is not known in advance, their resolution for accurate integration when calculating the above <<exotic>> schemes may take quite a long time, even when techniques accelerating the convergence of the Fourier-modal method are used.

One of the methods that make it possible to significantly reduce the number of evaluations of the scattering matrix when calculating spectra with narrow resonance lines is \edit{a} resonant mode approximation of the scattering matrix \cite {Gippius2005c, Bykov2013}. This method uses the fact that the spectral position of the resonant lines and the imaginary part of the corresponding eigenenergies are smooth functions of the lateral projections of the wavevector ${k_x, k_y}$. According to this approximation, \edit{the total scattering matrix can be represented as}
\begin{equation}
\mathbb{S}(\omega,k_x,k_y)=\mathbb{S}_b(\omega,k_x,k_y)+\sum_j\dfrac{|O_j\rangle\langle I_j|}{\omega-\omega_{r,j}},
\label{resapp1}
\end{equation}
where $\mathbb{S}_b$ is background scattering matrix, $|O_j\rangle$, and $\langle I_j|$ are the input and output ket and bra vectors obtained as a result of the procedure for finding the poles of the scattering matrix \cite{Gippius2005c}, $\omega_{r,j}$ is the complex resonant frequency, and $j$ is the pole number. Moreover, following the above, $\omega_{r,j}$, $|O_j\rangle$ и $\langle I_j|$ are smooth functions of the wavevector. The resonant mode approximation formulated in this way is well applicable when the frequency-wavevector range of interest is far from the opening of the diffraction channels. This limitation of the resonant mode approximation \eqref{resapp1} is due to the impossibility in it to take into account the multivalentness of the $\mathbb{S}$-matrix as a function of $\omega$. This leads to the fact that some of the poles of the $\mathbb{S}$-matrix are lost, and the expansion \edit{is no longer valid} when approaching the light cone. In practice, the dispersion curves of quasi-waveguide modes often intersect the air and (or) substrate light cones folded into the first Brillouin zone. Moreover, if it is necessary to integrate over the first Brillouin zone, we always pass the openings of the diffraction channels. In Refs.~\cite{Kirilenko1993, Lomakin2006, akimov2011optical} it was shown that it is much more convenient to use the expansion over the projection of the wavevector on the vertical $z$-axis, namely, $k_z$ for the accurate description of the optical properties 
in the resonant approximation when the dispersion curve of the resonance passes through the folded light cone. Instead of \eqref{resapp1} one should write (see Ref.~\cite{akimov2011optical})
\begin{equation}
\mathbb{S}(k_z,k_x,k_y)=\mathbb{S}_b(k_z,k_x,k_y)+\sum_{j}\dfrac{|O_j\rangle\langle I_j|}{k_z-k_{z,r,j}},
\label{resapp2}
\end{equation}
since the $\mathbb{S}$-matrix is a univalent function of $k_z$, which avoids the loss of a part of the poles. In the $k_z$-expansion, each quasiguided mode is described by two poles on the complex plane $\mathrm{Im}\,k_z/\mathrm{Re}\,k_z$. Near the diffraction threshold, both poles make approximately equal contributions to the scattering matrix, while far from it, one of the poles makes the predominant contribution.

Although the resonant mode approximation generalized in \cite{akimov2011optical} provides an exhaustive explanation of the importance and essence of the bipolar $k_z$-expansion, it also has a significant limitation. Namely, there must be no more than one light cone near the considered range of energies and wavevectors for its applicability. In this paper, we present a further generalization of the resonant mode approximation of the scattering matrix to the case of  {two} diffraction thresholds.

\section{Expression for the scattering matrix in the resonant mode approximation}
\edit{First, we briefly outline the main steps necessary for the construction of a scattering matrix. We consider an arbitrary \editd{multilayer periodic} structure, which can be divided into layers, each \editd{of which is}  homogeneous along the vertical $z$-axis and periodic or homogeneous along the lateral $x$- and $y$-axes. The periods of the structure along $x$- and $y$\editd{-axes} are denoted $d_{x}$ and $d_y$. Suppose, we are interested in the structure's optical response to an incident plane wave with a wavevector $\vec{k}=\{k_x,k_y,k_z\}$. Periodicity of the structure inevitably leads to the diffraction effect. The incident wave is transmitted and reflected to all the diffraction channels, i.e. to the waves with $k_x'=k_x+\dfrac{2\pi}{d_x}\alpha_x,\,k_y'=k_y+\dfrac{2\pi}{d_y}\alpha_y$\editd{, where} $\alpha=(\alpha_x,\alpha_y)$ is a diffraction order, \editd{with} $\alpha_{x,y}$ being integer numbers, $\dfrac{2\pi}{d_{x,y}}$ are the periods of the reciprocal lattice. To obtain a numerical solution to this problem, one has to choose some finite number $N_{x,y}$ of the plane waves or, namely, Fourier harmonics, that will be included in the numerical calculation.  Each Fourier harmonic is assigned a number $\alpha=(\alpha_x,\alpha_y)$, where $\alpha_x$ and $\alpha_y$ take values in the ranges from $-N_{x}$ to $N_{x}$ and from $-N_{y}$ to $N_{y}$ correspondingly. Further on, we will denote the diffraction order indices of the Fourier harmonics with the Greek letters. The total number of harmonics is then $N_g=(2N_x+1)\times(2N_y+1)$.
For each $\alpha$-th harmonic, there is a $z$-projection of the wavevector $k_z ^{(\alpha)}$ which in the case of a homogeneous medium with the dielectric permittivity $\epsilon$ can be found using the formula
\begin{equation}
    k_z^{(\alpha)}=\pm\sqrt{k^2\epsilon-\left(k_x-\frac{2\pi}{d_x}\alpha_x\right)^2-\left(k_y-\frac{2\pi}{d_y}\alpha_y\right)^2}.
    \label{kz_calc}
\end{equation}

Diffracted plane waves in the top and bottom outer layers  form a basis for the input and output states. In other words, the solution of Maxwell's equations outside of the structure is expanded into a sum of Fourier harmonics
\begin{equation}
\vec{E}(x, y, z, t)=\vec{E}^{(\alpha)} \exp \left(i\vec{k}^{(\alpha)}\vec{r}-i \omega t\right)
\label{FourierHarm}
\end{equation}
with the wavevectors 
\begin{equation}
    \vec{k}^{(\alpha)}=\{k_x-\frac{2\pi}{d_x}\alpha_x,k_y-\frac{2\pi}{d_y}\alpha_y, k_z^{(\alpha)}\}.
    \label{k_difr}
\end{equation} 
By definition \cite{Tikhodeev2002b}, the input vector consists of the amplitudes of the Fourier harmonics \eqref{FourierHarm} that propagate or decay towards the structure.  Amplitudes of those harmonics that propagate or decay away from the structure constitute the output vector. Finally, the scattering matrix of the structure connects the input and output vectors of amplitudes (see Fig.~\ref{ScatteringMatFig}):
\begin{equation}
    \underbrace{\left(\begin{array}{c}
\vec{\mathcal{A}}^{+}_b \\
\vec{\mathcal{A}}^{-}_t
\end{array}\right)}_{|Out\rangle}=\mathbb{S}(k_x,k_y,\omega)
\underbrace{\left(\begin{array}{c}
\vec{\mathcal{A}}^{+}_t \\
\vec{\mathcal{A}}^{-}_b
\end{array}\right)}_{|In\rangle}.
\label{Smat_def_append}
\end{equation}
Here $\vec{\mathcal{A}}^{+,-}_{t,b}$ denote the vectors of amplitudes that correspond to harmonics propagating \editd{or} decaying along and \editd{the positive and negative}
direction\editd{s} of $z$-axis (+ and - superscripts correspondingly) in the top and bottom outer layers (subscripts $t$ and $b$ correspondingly). Note that projections $k_x$, $k_y$ could generally take any values. Although for 2D periodic multilayers the knowledge of the scattering matrix within the first Brillouin zone is enough, this is not the case when it comes to 1D periodic structures or $xy$-homogeneous structures. This is because such structures do not have periodicity in the momentum space in one or two directions when only main harmonic along the corresponding direction is taken. Moreover, $k_x$, $k_y$ can be taken as complex numbers with positive imaginary parts responsible for decay along XY plane. 

\editd{As it is shown in Appendix A, the amplitude vector $\big(\vec{\mathcal{A}}^{+}, \vec{\mathcal{A}}^{-}\big)^\mathrm{T}$ in a certain $z$-coordinate represents a particular solution of Maxwell's equations determined by boundary conditions such as an incoming plane wave or absence of incoming waves.} For each diffraction order there are two directions of propagation, i.e. two signs in equation \eqref{kz_calc}, and two possible polarizations of light so that the output and input are vectors of length $4N_g$. For the sake of consistency, we provide the instructions of the scattering matrix calculation procedure in Appendix A.}

{The choice of the sign in \eqref{kz_calc} for the top and bottom outer layers affects the scattering matrix of the whole system (see the details in Appendix A). Indeed, according to the formalism of the Fourier modal method, the total scattering matrix of the whole system, $\mathbb{S}$, is found by the iterative procedure from the propagation matrices of the slabs, $\mathbb{P}$, and interface matrices of the interfaces between them, $\mathbb{I}$. The interface matrices are found as $\mathbb{I}_i = \mathbb{F}_{i-1}^{-1}\mathbb{F}_i$ where $\mathbb{F}_i$ the material matrix of the $i$-layer. By definition, a material matrix $\mathbb{F}$ connects the amplitudes of the optical modes propagating in the medium with the hypervectors of the Fourier components of the electric and magnetic fields. 
These eigenvectors as well as the propagation matrix $\mathbb{P}$ eventually depend on $k_z$ values and, hence, the total scattering matrix also depends on the $k_z$ values.}

{Thus, two scattering matrices calculated with different sign choices for $k_z$ of some Fourier harmonic are in fact different functions of energy. This fact was previously emphasized in works \cite{Kirilenko1993, akimov2011optical}}

When calculating the scattering matrix at real energies, the choice of the sign for each $k_z$ {in the substrate layer} is determined by the condition $\mathrm{Re}(k_z)>-\mathrm{Im}(k_z)$. At energies exceeding the energy of the diffraction threshold $\Omega_\alpha$, \editd{which can be} found as
\begin{equation}
\Omega_\alpha=\frac{\hbar c}{\sqrt{\epsilon}}\sqrt{\left(k_x-\frac{2\pi}{d_x}\alpha_x\right)^2+\left(k_y-\frac{2\pi}{d_y}\alpha_y\right)^2},
\end{equation}
the $z$-projection of the wavevector of the $\alpha$-th order Fourier harmonic {calculated using \eqref{kz_calc}} is a positive real number {$\mathrm{Im}(k_z^{(\alpha)})=0,\,\mathrm{Re}(k_z^{(\alpha)})>0$}, while at energies below the threshold, it is a purely imaginary number {$\mathrm{Im}(k_z^{(\alpha)})>0,\,\mathrm{Re}(k_z^{(\alpha)})=0$}. {Such convention guarantees that we distinguish outgoing and incoming waves. Waves in the substrate depend on $z$ coordinate as $\exp(ik_z z)$ and are either propagating or exponentially decreasing along the positive direction of $z$-axis. These waves are outgoing waves. At the same time, incoming waves in the substrate layer are described by $\exp(-ik_z z)$ as they propagate or decay towards the structure, in the direction opposite to the direction of the $z$-axis.} Thus, with an increase of the photon energy, physically meaningful values of $k^{(\alpha)}_z$ {follow} the trajectory {of a right angle} $\mathcal{C}_{k_z}$ on a complex $k_z$ plane 
\begin{equation}
\mathcal{C}_{k_z}=\{\mathrm{Im}(k_z^{(\alpha)})=\infty \rightarrow k_z^{(\alpha)}=0 \rightarrow \mathrm{Re}(k_z^{(\alpha)})=\infty\},
\end{equation}
with the angle vertex $k_z^{(\alpha)}=0$ corresponding to the energy $\Omega_n$.

In finding the poles of the scattering matrix, it is necessary to calculate the scattering matrix at complex energies. At each iteration of the computational algorithm (see Ref.~\cite{Gippius2010} or Appendices A,B for the details), it is necessary to choose the sign before the root in expression \eqref{kz_calc}. {As mentioned above, } the total scattering matrix of the overall system essentially depends on the choice of the sign of each $k^{(\alpha)}_z$. \edit{Two matrix functions $\mathbb{S}(\omega, k_x, k_y, k_z ^{(\alpha)})$ and $\mathbb{S}(\omega, k_x, k_y, -k_z ^{(\alpha)})$ \edit{that} differ only in the choice of the sign of the root in expression \eqref{kz_calc} 
are generally different functions of energy. Thus they may have resonant poles at different energy.} 

Nevertheless, far from the diffraction thresholds, the selection rule $\mathrm{Re}(k_z)>-\mathrm{Im}(k_z)$ {can be} successfully applied. Suppose that we are searching for the resonances in the region near the real energy $E=E_0$, which corresponds to the value of $k_{z, 0}^{(\alpha)}=\sqrt{\left(\dfrac{E}{\hbar c}\right)^2\epsilon-\left(k_x-\frac{2\pi}{d_x}\alpha_x\right)^2-\left(k_y-\frac{2\pi}{d_y}\alpha_y\right)^2}$. {Here $n$ denotes the index number of the Fourier harmonic, such that the corresponding diffraction threshold $\Omega_\alpha$ appears to be the closest to $E_0$. For all other harmonics the following discussion is valid as well.} When calculating the scattering matrix at an energy $E = E_0-i \Gamma$, $\Gamma\ll E_o-\Omega_\alpha$, the $z$-projection of the wave vector of the $\alpha$-th harmonic is
\begin{gather}
    k_z^{(\alpha)}=\pm\dfrac{\sqrt{\epsilon}}{\hbar c}\sqrt{E^2-\Omega_\alpha^2}\approx\pm\dfrac{\sqrt{\epsilon}}{\hbar c}\sqrt{E_0^2-2i\Gamma E-\Omega_\alpha^2}\approx\nonumber\\
    \approx\pm k_{z,0}^{(\alpha)}(1-\dfrac{i\Gamma E_0}{E_0^2-\Omega_\alpha^2})
\end{gather}

We will consider now two possible scenarios. First, suppose we found a pole at the point $E = E_0-i \Gamma, \, k_z^{(\alpha)} {\approx} + k_{z,0}^ {(\alpha)} (1-\dfrac{i \Gamma E_0}{E_0^2-\Omega_\alpha^2})$, i.e. in some energy range near $\omega_0=E_0/\hbar$ one can use \eqref{resapp1} with the only pole $\omega_r=(E_0-i \Gamma)/\hbar$ and $\mathbb{S}(\omega)=const$. \edit{The pole-search algorithm is was presented in Ref.~\cite{Gippius2010} and is additionally described in Appendix B.} In the resonant mode approximation {\eqref{resapp1}}, the contribution $S_r$ of this pole at the point $(E_0,\, k_{z, 0}^{(\alpha)})$ in expansion in the energy space will be 
\begin{equation}
    S_r^{(E)}(E_0)=\dfrac{|O^{(E)}\rangle\langle I^{(E)}|}{E_0-(E_0-i\Gamma)}=\dfrac{S_r^{(E)}}{i\Gamma},
    \label{Resonance_energy}
\end{equation} and in expansion in the $k_z^{(\alpha)}$ space: 
\begin{equation}
    S_r^{(\alpha)}(k_{z,0}^{(\alpha)}){\approx}\dfrac{|O^{(\alpha)}\rangle\langle I^{(\alpha)}|}{\dfrac{i\Gamma E_0}{\sqrt{E_0^2-\Omega_\alpha^2}}\dfrac{\sqrt{\epsilon}}{\hbar c}}.
\end{equation}
{Expression \eqref{Resonance_energy} represents the fact that the total scattering matrix found by the FMM iteration procedure may have the pole singularity. Irrespective a physical nature of this pole singularity we represent the scattering matrix in the form \eqref{Resonance_energy}}.

One can show the equivalence of these two approximations. Suppose we have a scattering matrix function $\mathbb{S}(E)$ with a pole in $E_r$, and there are two variables $E$ and $E'=E'(E)$, so that $\dfrac{dE'}{dE}|_{E=E_r}=C_1$. Then for $E$ close enough to $E_r$ one can write a resonant mode  approximation for $\mathbb{S}$ as a function of $E$ and also as a function of $E'$.
{
\begin{equation}
\mathbb{S}(E)=\mathbb{S}_b(E)+\dfrac{|O\rangle\langle I|}{E-E_r}=\mathbb{S}_b(E')+\dfrac{|O'\rangle\langle I'|}{E'-E'_r}=\mathbb{S}_b(E)+\dfrac{|O\rangle\langle I|C_1}{C_1 E-C_1 E_r}.
\end{equation}

Let us consider this equation near the resonant pole $E=E_r$, where $\mathbb{S}_b(E)\approx\mathbb{S}_b(E')\approx const$.} One can see that the outer products of the resonant out- and in-vectors in $E$ and $E'$ expansions are equivalent up to a scaling factor:
\begin{equation}
    |O'\rangle\langle I'|=|O\rangle\langle I|\dfrac{dE'}{dE}(E=E_r).
    \label{Pole_equivalence}
\end{equation}

Taking into account the expression
\begin{equation}
    \dfrac{dk_z^{(\alpha)}}{dE}=\dfrac{\sqrt{\epsilon}}{\hbar c}\dfrac{E}{\sqrt{E^2-\Omega_\alpha^2}}
\end{equation}
and \eqref{Pole_equivalence} we obtain
\begin{equation}
    S_r^{(E)}(E_0)\approx S_r^{(\alpha)}(k_{z,0}^{(\alpha)}).
\end{equation}

Now let us examine the second case, when the pole is at the point $E=E_0-i\Gamma,\,k_z^{(\alpha)}=-k_{z,0}^{(\alpha)}(1-\dfrac{i\Gamma E_0}{E_0^2-\Omega_\alpha^2})$. The contribution of this pole at any point on the trajectory of real energies $\mathcal{C}_{k_z}$ in the expansion in the $k_z^{(\alpha)}$-space will be at most
\begin{equation}
    \|S_r^{(\alpha)}(k_{z}^{(\alpha)} \in {C}_{k_z})\|\lessapprox\|\dfrac{|O^{(k_z^{(\alpha)})}\rangle\langle I^{(k_z^{(\alpha)})}|}{k_{z,0}^{(\alpha)}}\|.
\end{equation}
Such a pole turns out to be very distant from the trajectory of real energies $\mathcal{C}_{k_z}$, and therefore its influence is negligible. The standard condition for choosing $\mathrm{Re}(k_z)>-\mathrm{Im}(k_z)$ guarantees that such a pole will not be found. We also note that, in contrast to the first case, for this pole, the resonant mode approximation in the energy space is entirely inapplicable, since it does not distinguish between the first and second poles, and, therefore, gives an inappropriately enormous contribution \eqref{Resonance_energy} at the point $E_0, k_{z, 0}$ in the second case.

{The situation changes significantly when in the region of interest the condition $\Gamma \ll E_o-\Omega_\alpha$ ceases to be fulfilled}. In this case, both poles are of equal importance, and the resonant mode approximation in energy space is not applicable. A detailed study of this issue is given in Ref.\,\cite{akimov2011optical}. We emphasize once again that the poles $\mathrm{Re}(k_z)<-\mathrm{Im}(k_z)$ are neither incorrect nor non-physical; it is just that their search and consideration are expedient only near diffraction thresholds.

When two diffraction thresholds are present in the energy/wavevector range of interest, the resonant mode approximation of the scattering matrix as a function of only one $k_z$ is not sufficient anymore. The reason is that in the optical spectra of the structure, two distinctive root features appear at energies corresponding to the opening of the diffraction channels. These features cannot be simultaneously reproduced in the resonant mode approximation in the space of one of $k_z$, since the trajectory $\mathcal{C}_{k_z}$ is not smooth at only one point.

Let us deduce a new resonant mode approximation that takes into account both $k_z^{(\alpha)}$ and $k_z^{(\beta)}$ dependencies, with $\alpha$ and $\beta$ denoting the indices of the Fourier harmonics, such that the diffraction thresholds $\Omega_{\alpha,\beta}$ appear within the energy range of interest.
{From \eqref{kz_calc} we obtain:
\begin{equation}
    {k^{(\alpha)}_{z}}^2-{k^{(\beta)}_{z}}^2=\dfrac{4\pi^2}{d_x^2}(\alpha_x^2-\beta_x^2)-\dfrac{4\pi k_x}{d_x}(\alpha_x-\beta_x)+\dfrac{4\pi^2}{d_y^2}(\alpha_y^2-\beta_y^2)-\dfrac{4\pi k_y}{d_y}(\alpha_y-\beta_y)
\end{equation}
For a selected angle of incidence, i.e. $k_{x,y}=const$}:
\begin{equation}
d({k^{(\alpha)}_{z}}^2-{k^{(\beta)}_{z}}^2)=0,
\end{equation}
which means, that in the vicinity of {some scattering matrix pole $k^{(\alpha)}_{z,r}$} the following relation is valid:
\begin{equation}
    dk^{(\alpha)}_{z}=dk^{(\beta)}_{z}\dfrac{k^{(\beta)}_{z,r}}{k^{(\alpha)}_{z,r}}.
    \label{dkz_relation}
\end{equation}
According to the \eqref{Pole_equivalence} and \eqref{dkz_relation} we derive:
\begin{equation}
    |O^{(\beta)}\rangle\langle I^{(\beta)}|=|O^{(\alpha)}\rangle\langle I^{(\alpha)}|\dfrac{k^{(\alpha)}_{z,r}}{k^{(\beta)}_{z,r}}.
\label{ResMat_Kz_Scale}
\end{equation}

 In the vicinity of the pole, the new resonant mode approximation, which accounts for the pole trajectory in the space of two variables $k^{(\alpha)}_z$ and $k^{(\beta)}_z$ {simultaneously}, should also be equivalent to the previous realization of the resonant mode approximation \eqref{resapp2} in  $k^{(\alpha)}_z$ and $k^{(\beta)}_z$ spaces separately:
 \begin{gather}
     \dfrac{|O^{(\alpha)}\rangle\langle I^{(\alpha)}|}{k^{(\alpha)}_z-k^{(\alpha)}_{z,r}}{\approx}\dfrac{|O^{(\beta)}\rangle\langle I^{(\beta)}|}{k^{(\beta)}_z-k^{(\beta)}_{z,r}}{\approx}
      \dfrac{|O^{(\alpha;\beta)}\rangle\langle I^{(\alpha;\beta)}|}{\rho(k^{(\alpha)}_z,k^{(\beta)}_z,k^{(\alpha)}_{z,r},k^{(\beta)}_{z,r})}.
      \label{approx_equiv}
 \end{gather}
{We choose the new resonant denominator with the simplest possible structure 
\begin{equation}
    \rho(k^{(\alpha)}_z,k^{(\beta)}_z,k^{(\alpha)}_{z,r},k^{(\beta)}_{z,r})=k^{(\alpha)}_{z}-k^{(\alpha)}_{z, r} + k^{(\beta)}_{z}-k^{(\beta)}_{z, r},
\end{equation} which has two points with the discontinuity of derivatives as a function of energy at 
$E=\Omega_{\alpha,\beta}$. 
Applying equation \eqref{dkz_relation} in the vicinity of the resonant pole we obtain:
\begin{equation}
    \rho(k^{(\alpha)}_z,k^{(\beta)}_z,k^{(\alpha)}_{z,r},k^{(\beta)}_{z,r})\approx \left(k^{(\alpha)}_{z}-k^{(\alpha)}_{z, r}\right)\left(1+\dfrac{k^{(\alpha)}_{z,r}}{k^{(\beta)}_{z,r}}\right).
\end{equation}
Then, in order to satisfy equivalence of approximations 
\eqref{approx_equiv}}: 
\begin{multline}
  |O^{(\alpha;\beta)}\rangle\langle I^{(\alpha;\beta)}|=|O^{(\alpha)}\rangle\langle I^{(\alpha)}|\left(1+\dfrac{k^{(\alpha)}_{z,r}}{k^{(\beta)}_{z,r}}\right)=\\=  |O^{(\alpha)}\rangle\langle I^{(\alpha)}|+|O^{(\beta)}\rangle\langle I^{(\beta)}|.
\end{multline}
 Hence, the new resonant mode approximation takes the form:
\begin{equation}
    \mathbb{S}(\omega,k_x,k_y)=\mathbb{S}_b(\omega,k_x,k_y)+\sum_n\dfrac{|O^{(\alpha)}_n\rangle\langle I^{(\alpha)}_n|+|O^{(\beta)}_n\rangle\langle I^{(\beta)}_n|}{k^{(\alpha)}_{z}-k^{(\alpha)}_{z, r,n} + k^{(\beta)}_{z}-k^{(\beta)}_{z, r,n}},
    \label{NewResApprox}
\end{equation}
where $n$ denotes the pole index number, $\alpha$ and $\beta$ are the indices of the Fourier harmonics which diffraction occurs within the energy range of interest. In the case of several diffraction thresholds $\alpha={\alpha_1,\alpha_2,...}$ in the approximation region we write a general equation:
\begin{equation}
    \mathbb{S}(\omega,k_x,k_y)=\mathbb{S}_b(\omega,k_x,k_y)+\sum_n\dfrac{\sum_{\alpha}|O^{(\alpha)}_n\rangle\langle I^{(\alpha)}_n|}{\sum_\alpha\left(k^{(\alpha)}_{z}-k^{(\alpha)}_{z,r,n}\right)}.
    \label{GeneralResApprox}
\end{equation}
Please note that the new resonant approximation \eqref{NewResApprox} and \eqref{GeneralResApprox} does not require calculating $|O^{(\alpha)}_n\rangle\langle I^{(\alpha)}_n|$ for all harmonics, because, according to \eqref{ResMat_Kz_Scale}:
\begin{equation}
    \sum_\alpha|O^{(\alpha)}_n\rangle\langle I^{(\alpha)}_n|=|O^{(\beta)}_n\rangle\langle I^{(\beta)}_n|\left(1+\sum_{\alpha\ne \beta}\dfrac{k^{(\beta)}_{z,r,n}}{k^{(\alpha)}_{z,r,n}}\right),
\end{equation}
where $\beta$ is an index of any Fourier harmonic that we use for the resonant mode approximation construction.

\section{Numerical example}
\begin{figure}
\begin{center}
\includegraphics[width=0.6\linewidth]{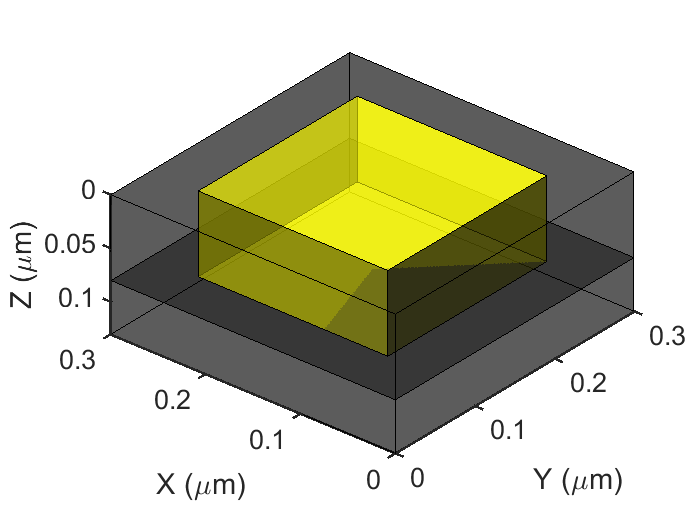}
\caption{Periodicity cell of the structure of interest.}
\label{Structure}
\end{center}
\end{figure}

For numerical verification of the new resonant mode approximation, we use a 2D periodic structure, consisting of an 80-nm thick quartz ($\epsilon_{SiO_2}=2.25$) layer with an embedded square lattice of rectangular ZnO square inclusions ($\epsilon_{ZnO}=6.25$) sandwiched between the air and the substrate as top and bottom semi-infinite layers (see Fig. \ref{Structure}). The structure period is $d=300$~nm, and the ZnO square side is $w=200$~nm.

\begin{figure*}
    \centering
    \includegraphics[width=1\linewidth]{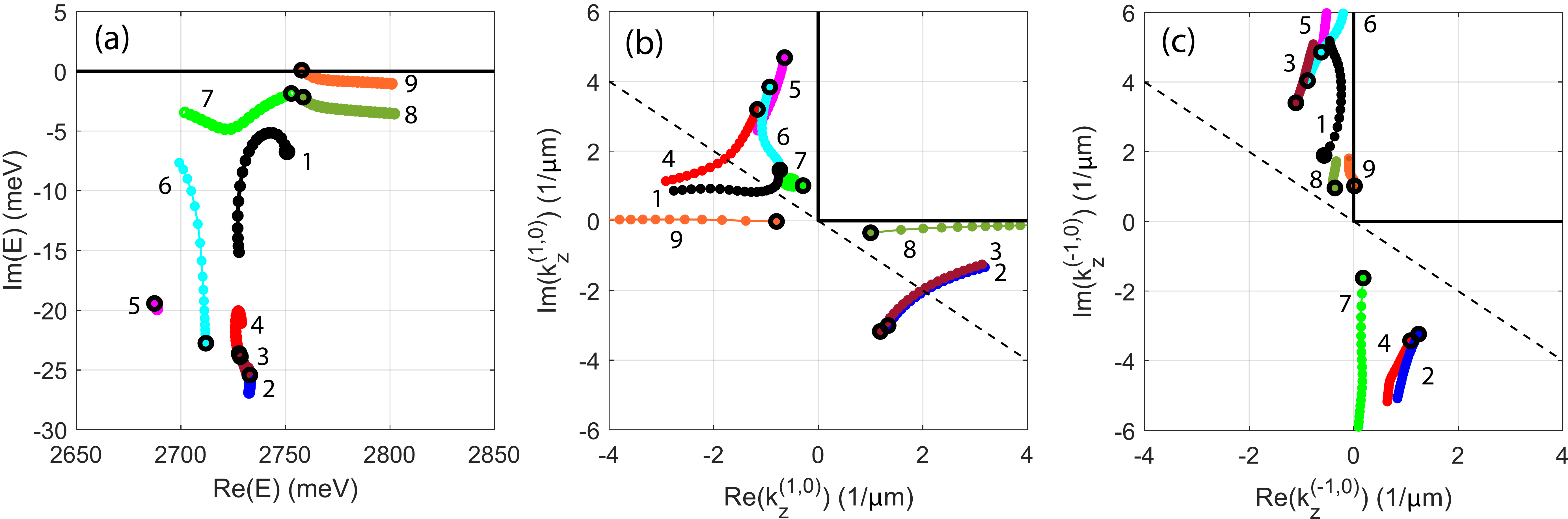}
    \caption{Calculated trajectories of the resonant (a) energies, (b) $k_z^{(1,0)}$, (c) $k_z^{(-1,0)}$. $k_x $ changes from 0.02 to 0.4 $\mu$m$^{-1} $. Each resonance has the same color on all three panels. Black empty circles indicate the positions of the resonances at $ k_x = $ 0.02 $\mu$m$^{-1} $. Thick black lines indicate the real energy trajectory $\mathcal{C}_{k_z}$, thin dashed lines are described by the equation $\mathrm{Re}(k_z)<-\mathrm{Im}(k_z)$. }
    \label{fig:2}
\end{figure*}

\begin{figure*}
    \centering
    \includegraphics[width=1\linewidth]{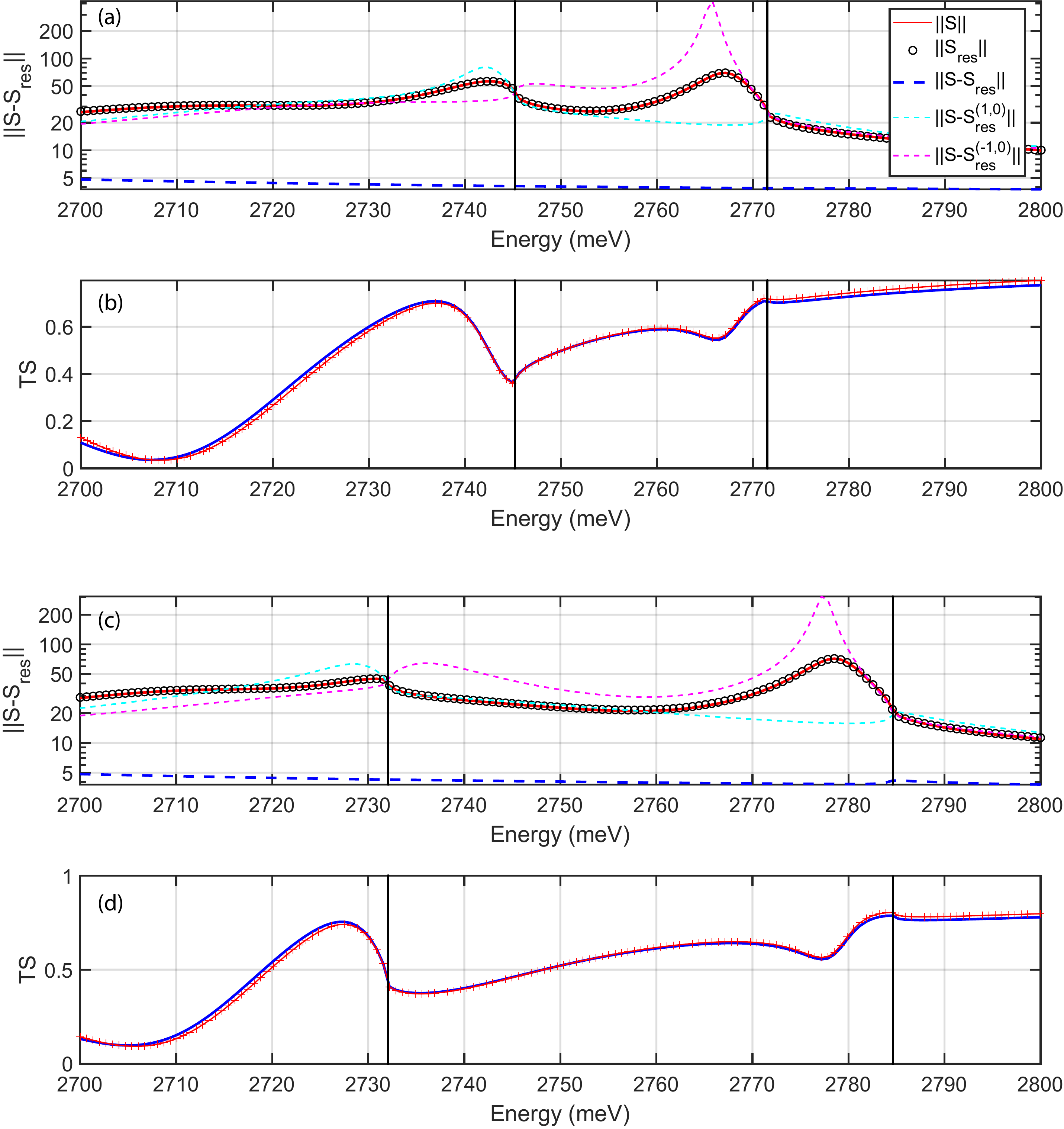}
    \caption{Comparison of the exact calculation and the resonant mode  approximation for $k_y = 1$\,$\mu$m $^{-1}$ and (a, b) $k_x = 0.1$\,$\mu$m $^{-1}$ and (c, d) $k_x = 0.2$\,$\mu$m $^{-1}$. On the top panels, the spectral norm of the exactly calculated scattering matrix is indicated by a continuous red line, the spectral norm of the resonant matrix without taking into account $\mathbb{S}_b$ is indicated by black circles, the spectral norms of the difference between the exact matrix and the resonant mode approximation are shown by blue, cyan, and purple dashed lines for resonant mode approximations simultaneously in $k_z^{(\pm 1, 0)}$, only for $k_z^{(1, 0)}$, only for $k_z^{(-1, 0)}$, respectively. Black vertical lines correspond to diffraction thresholds. The lower panels show the dependencies of the transmittance of the structure in $s$-polarization. The exact calculation is shown with a solid blue line, the result of the resonant mode approximation is shown with red crosses.}
    \label{Spectra}
\end{figure*}

We intend to calculate the spectral norm of the scattering matrix and the transmission spectra of the above structure at photon energies within the range between 2700~meV and 2800~meV at $k_y=1\,\mu$m$^{-1}$ using the standard Fourier modal method as well as the developed resonant mode approximation. For this, we rigorously found all significant resonances in this energy range for $k_x=0.02\mu$m$^{-1}$. Then we consequentially traced their trajectories with $k_x$ varied between $k_x=0.02\mu$m$^{-1}$ and $k_x=0.4\mu$m$^{-1}$. Such selection of parameters guarantees the sufficiency of the resonant mode approximation with only two $k_z$ corresponding to $2N_x=\pm 1,\,N_y = 0$ harmonics {in the substrate layer}. The resonant mode approximation in the proximity of more than two diffraction channels is also possible but is much more difficult for the presentation. Indeed, as it was shown in Ref.\,\cite{akimov2011optical}, one resonant peak in the vicinity of $\alpha$-th harmonic diffraction threshold requires two $k^{(\alpha)}_z$ poles for successful reproduction of the spectrum, while all other $k_z$ are chosen according to the rule $\mathrm{Re}(k_z)>-\mathrm{Im}(k_z)$. Now, in the vicinity of  {two} diffraction thresholds (corresponding to $\alpha$-th and $\beta$-th harmonics) we also should consider additional poles that satisfy the condition $\mathrm{Re}(k^{(\beta)}_z)<-\mathrm{Im}(k^{(\beta)}_z)$; hence, the number of significant poles is approximately doubled. Moreover, including a new diffracting harmonic into the resonant mode approximation always doubles the number of poles. Although finding a pole is a rather routine procedure itself, the entire set of poles makes the application of the resonant mode approximation fairly complicated.

In the demonstrated example we found nine poles shown in Fig.\,\ref{fig:2}. Fig.\,\ref{fig:2}(a) represents resonant energies, while Figs.\,\ref{fig:2}(b) and (c) show the trajectories of the resonant modes in $k^{(\pm 1,0)}_z$ spaces. Here we emphasize, that the general pole search procedure in energy space is only capable to detect poles 1,\,5,\,6, and 8. Pole 1 disappears as $k_x$ shifts to larger values, because $k^{(1,0)}_{z,r}(k_x)$ inevitably crosses the dashed line $\mathrm{Re}(k^{(1,0)}_z)=-\mathrm{Im}(k^{(1,0)}_z)$. We also notice that the resonant energy and resonant $k^{(-1,0)}_{z,r}$ of pole 9 at $k_x=0.02\mu$m$^{-1}$ lie approximately on the real energy trajectory. This causes almost an infinite growth of this pole contribution when using previous realization of the resonant mode approximations either in the energy space \eqref{resapp1} or in the single-$k_z$ space \eqref{resapp2}. In Fig.\,\ref{Spectra} it can be seen that the new realization of the resonant mode approximation solves this problem. 

Fig.\,\ref{Spectra} shows the comparison of the exact calculations made by the Fourier modal method ($2N_{x}+1=11$, $2N_{y}+1=11$) with the results of resonant mode approximation \eqref{NewResApprox}. One can see that the resonant mode approximation perfectly reproduces all the peculiarities of the optical spectra, including unsmoothnesses at the energies of the diffraction thresholds. Additionally, the single-$k_z$ resonant mode approximation is not valid in the vicinity of two diffraction thresholds.

{ Positions of resonances, as well as the resonant output and input vectors, are smooth functions of $k_{x,y}$. Thus, for retrieving a $\mathbb{S}(\omega, k_x, k_y)$ dependence in a broad energy/wavevector range we suggest first finding the resonant poles and vectors for some selected parameters $k_{x,y}$ and later using these resonant energies as initial guesses in the pole finding procedure. One can consequently calculate $\mathbb{S}(\omega)$ in the resonant approximation for all values of $k_{x,y}$ in a range of interest by tracing resonant energies as as presented in Fig. 2. This also appears to be a computationally efficient approach as one does not have to calculate optical spectra directly for each given $k_{x,y}$.
}

\section{Conclusions}
In conclusion, we developed the resonant mode approximation applicable to systems with two diffraction threshold openings at the energy/wavevector range of interest. We have demonstrated that the resonant mode approximation perfectly reproduces the scattering matrix norm and the transmission spectra calculated with a standard Fourier modal method in the energy range where the previous realizations of this approximation are not applicable. While the developed technique opens up the opportunity to significantly accelerate the scattering matrix calculation rate, a large number of poles may complicate its practical implementation.

\section*{Acknowledgments}
This work was supported by the Russian Science Foundation (project \textnumero 21-72-20184)

\appendix
\edit{
\section{Scattering matrix formalism}
For the sake of completeness of this paper, we provide the scattering matrix calculation algorithm following after Ref.~\cite{Tikhodeev2002b}. Derivation of the scattering matrix of a periodic structure is performed in 3 following steps.

1. The solution of Maxwell's equations in each layer of the structure is decomposed into a sum of plane waves of the form
\begin{equation}
\vec{E}(x, y, z, t)=\sum_{(\alpha)}\vec{E}^{(\alpha)} \exp \left[i(k_x-\frac{2\pi}{d_x}\alpha_x)x+i(k_y-\frac{2\pi}{d_y}\alpha_y)y\right]\exp(iKz-i \omega t).
\label{FourierSum}
\end{equation}
Note that $K$ here is an unknown $z$-projection of a wavevector that corresponds to some mode of the periodic layer. $K$ could be calculated analytically only for the case of a homogeneous layer using equation \eqref{kz_calc} when each mode of the layer consists of only one Fourier harmonic. Substituting sum \eqref{FourierSum} into Maxwell's equations and excluding the $z$-projections of the electric field one arrives at an eigenproblem
\begin{equation}
    \mathcal{M}\vec{\mathcal{E}}_{||}=\vec{\mathcal{E}}_{||}K^2.
    \label{MasterEq}
\end{equation}
Eigenvalues here are  propagation constants $K$, and the corresponding eigenvectors \begin{equation}
   \vec{\mathcal{E}}_{||}=\left[E_{x}^{(\alpha_1)},...,E_{x}^{(\alpha_{N_g})},E_{y}^{(\alpha_1)},...,E_{y}^{(\alpha_{N_g})}\right]^T 
\end{equation} are vectors composed of the lateral projections of the electric fields of the Fourier harmonics from equation \eqref{FourierSum} that constitute this particular mode of the layer. For a detailed derivation of the matrix $\mathcal{M}$, an engaged reader could address Ref.~\cite{Tikhodeev2002b}.

2. Once we have derived the optical modes of each slab that propagate along \editd{the positive and negative $z$-directions} 
as $\exp(\pm iKz)$, we combine them to satisfy the boundary conditions. Given a specific set of eigenvectors 
\begin{equation}
    \mathcal{E}=\left(\vec{\mathcal{E}}_{||}^{(1)},\vec{\mathcal{E}}_{||}^{(2)},...,\vec{\mathcal{E}}_{||}^{(2N_g)}\right)
    \label{MatrixE}
\end{equation}
in the $n$-th layer (superscript $i$ denotes the serial number of the eigenvector), the actual electromagnetic field would be a linear combination of such layer modes with amplitudes $\vec{\mathcal{A}}^{\pm}$ for positive and negative propagation direction:
\begin{equation}
\vec{\mathrm{A}}(z)=\left(\begin{array}{c}
\vec{\mathcal{A}}^{+}(z) \\
\vec{\mathcal{A}}^{-}(z)
\end{array}\right).
\end{equation}
Inside one particular layer, each mode has a defined propagation constant. \editd{Connecting} the amplitudes of the modes at two different vertical coordinates $z$ and $z'$ we write:
\begin{equation}
\vec{\mathrm{A}}(z')=\mathbb{P}_{z'-z} \vec{\mathrm{A}}(z)
\end{equation}
    \quad \begin{equation}
\mathbb{P}_{L}=\left(\begin{array}{cc}
\exp [i \mathcal{K} (z'-z)] & 0 \\
0 & \exp [-i \mathcal{K} (z'-z)]
\end{array}\right),
\end{equation}
where matrix $\mathcal{K}$ is a diagonal matrix composed of propagation constants $K$ for the selected layer. $\mathbb{P}_{L}$ is called \editd{a} propagation matrix. Lateral projections of the electric and magnetic fields should be continuous on the interfaces between two layers. The electric and magnetic fields are calculated using the concept of the material matrix. By definition, a material matrix $\mathbb{F}$ connects the amplitudes of the optical modes propagating in the medium with the vectors of the {Fourier components} of the electric $\vec{\mathcal{E}}_{||}$ and magnetic $\vec{\mathcal{H}}_{||}$ fields. The material matrix has a block form:
\begin{equation}
\mathbb{F}=\begin{pmatrix}
\mathcal{E}&\mathcal{E}\\
\mathcal{H}&-\mathcal{H}
\end{pmatrix}, \quad 
\left(\begin{array}{l}
\vec{\mathcal{E}}_{||}(z) \\
\vec{\mathcal{H}}_{||}(z)
\end{array}\right)=\mathbb{F}\vec{\mathrm{A}}(z)
\label{Material}
\end{equation}
where $\mathcal{E}$ and $\mathcal{H}$ are $2N_g\times2N_g$ matrices composed of the {layer} eigenvectors from \eqref{MasterEq} as defined in \eqref{MatrixE} (see Ref.\,\cite{Tikhodeev2002b} for the details and rigorous derivation of $\mathcal{H}$). Taking into account the boundary conditions on the interface at a coordinate $z_n$ between the $n$-th and the $n+1$-th layers, we derive:
\begin{equation}
    \left(\begin{array}{l}
\vec{\mathcal{E}}_{||}(z_n-0) \\
\vec{\mathcal{H}}_{||}(z_n-0)
\end{array}\right)=
    \left(\begin{array}{l}
\vec{\mathcal{E}}_{||}(z_n+0) \\
\vec{\mathcal{H}}_{||}(z_n+0)
\end{array}\right), \quad \vec{\mathrm{A}}(z_n+0)=\mathbb{F}_{n+1}^{-1}\mathbb{F}_n\vec{\mathrm{A}}(z_n-0).
\end{equation}
Here the product of the inverse material matrix of the $n+1$-th layer and the material matrix of the $n$-th layer is called an interface matrix. Iteratively combining propagation and interface matrices one can construct the transfer matrix of the whole system $\mathbb{T}$, \editd{which} connects the amplitudes of the Fourier harmonics infinitesimally above the top boundary  of the structure $\vec{\mathcal{A}}^{\pm}_t$ with the amplitudes infinitesimally below the bottom boundary $\vec{\mathcal{A}}^{\pm}_b$ :
\begin{equation}
    \left(\begin{array}{c}
\vec{\mathcal{A}}^{+}_b \\
\vec{\mathcal{A}}^{-}_b
\end{array}\right)=\mathbb{T}
\left(\begin{array}{c}
\vec{\mathcal{A}}^{+}_t \\
\vec{\mathcal{A}}^{-}_t
\end{array}\right)
\label{Transfer}
\end{equation}

3. Finally, we rearrange the amplitudes to distinguish the incoming and outgoing waves. The scattering matrix $\mathbb{S}$ is obtained from the transfer matrix $\mathbb{T}$ \eqref{Transfer} in accordance with definition \eqref{Smat_def_append}:
\begin{equation}
    \mathbb{S}=\begin{pmatrix}
    \mathbb{T}_{11}-\mathbb{T}_{12}\mathbb{T}_{22}^{-1}\mathbb{T}_{21}& \mathbb{T}_{12}\mathbb{T}_{22}^{-1}\\-\mathbb{T}_{22}^{-1}\mathbb{T}_{21}& \mathbb{T}_{22}^{-1}
    \end{pmatrix}
\end{equation}
However, for numerical stability of the calculation, it is better to iteratively combine scattering matrices of the layers instead of the transfer matrices (propagation and interface matrices). Such an approach prevents the multiplication of the exponential factors produced by the evanescent waves $\sim\exp(-iKz)$ with $\mathrm{Im}(K)<0$.

\section{Resonant mode approximation for the scattering matrix and the pole-search procedure}
For consistency of the manuscript, here we also explain how optical eigenmodes are derived from the scattering matrix and provide the
resonant mode approximation formalism following after Ref.~\cite{Gippius2010}.
Optical modes are nontrivial solutions of Maxwell's equations with no incoming waves \cite{Gippius2005c, Lalanne2019}. They are characterized by complex energies $\omega=\Omega-i\Gamma$, so that the resonant field amplitudes decay in time as $|\exp(-i\omega t)|= \exp(-\Gamma t)$. Resonant modes, i.e. solutions of equation \eqref{Smat_def_append} with $\left|In\right\rangle=0$, could be found using the linearization of the inverse scattering matrix:
\begin{equation}
   \mathbb{S}^{-1}(\omega,k_{||})\left|O\right\rangle=0.
   \label{PoleDef}
\end{equation}
Here $k_{||}=\{k_x,k_y\}$ is a fixed projection of the wavevector, $\omega$ is a complex energy of an optical eigenmode, and $|O\rangle$ is the resonant output vector consisting of the Bragg harmonics complex amplitudes that constitute the resonant mode. By expanding the inverse scattering matrix in the Taylor series up to the linear term in energy, we write:
\begin{equation}
\mathbb{S}^{-1}(\omega)|O\rangle \approx\left[\mathbb{S}^{-1}\left(\omega'\right)+\left.\left(\omega-\omega'\right) \dfrac{\partial \mathbb{S}^{-1}}{\partial \omega}\right|_{\omega'}\right]|O\rangle = 0.
\label{Iterative}
\end{equation}
Equation \eqref{Iterative} \editd{has to be} solved iteratively. Given an initial guess value of energy $\omega'$ one should calculate $\mathbb{S}(\omega')$ and $\left.\dfrac{\partial\mathbb{S}^{-1}}{\partial \omega}\right|_{\omega'}$ to solve the eigenproblem
\begin{equation}
    -\left(\left.\dfrac{\partial\mathbb{S}^{-1}}{\partial \omega}\right|_{\omega'}\right)^{-1}\mathbb{S}(\omega')\mathrm{X}=\mathrm{X}\Delta
    \label{EigenIteration}
\end{equation}
that follows from the right side of equation \eqref{Iterative}. Here $\mathrm{X}$ is a square matrix which columns are the eigenvectors and $\Delta=\text{diag}\{\delta_i\}$ is a diagonal matrix of the eigenvalues. Taking the smallest in modulus eigenvalue $\delta_1$ as an iteration step, we obtain $\omega=\omega'+\delta_1$. The iterative procedure is then repeated until the sufficient precision in resonant energy determination is reached.

To derive the resonant approximation of the scattering matrix in the vicinity of a pole we extract $\mathbb{S}^{-1}(\omega')$ from equation \eqref{EigenIteration} and substitute the result into equation \eqref{Iterative}. By taking the inverse, we deduce:
\begin{equation}
\mathbb{S}(\omega) \approx \mathrm{X}\left[\left(\omega-\omega'\right) \mathbb{I}-\Delta\right]^{-1} \mathrm{Y}=\mathbb{S}_b+\sum_{n=1}^{N}\left|O_{n}\right\rangle \frac{1}{\omega-\omega_{n}}\left\langle I_{n}\right|,
\end{equation}
where $\mathbb{I}$ is the identity matrix, $\mathbb{S}_b$ denotes a non-resonant background scattering matrix. The output and input resonant vectors $|O\rangle$, $\langle I|$ are defined as columns and rows of matrices $\mathrm{X}$ and $\mathrm{Y}$ correspondingly:

\begin{equation}
\mathrm{X}=\left(\begin{array}{lll}
\left|O_{1}\right\rangle, & \left|O_{2}\right\rangle, & \ldots
\end{array}\right), \quad \mathrm{Y}=
\left(\left.\dfrac{\partial \mathbb{S}^{-1}}{\partial \omega}\right|_{\omega'}
\mathrm{X}
\right)^{-1}=\left(\begin{array}{c}
\left\langle I_{1}\right| \\
\left\langle I_{2}\right| \\
\vdots
\end{array}\right).
\end{equation}

It turns out that consideration of all poles $\omega'+\delta_i$, calculated at some energy $\omega'$ is not practical. A root  of linearized equation \eqref{Iterative} tend\editd{s} to a true solution of the equation \eqref{PoleDef} only if the linearization \eqref{Iterative} is valid for $\omega=\omega'+\delta_i$. Moreover, consideration of all the distant resonances seems to be redundant, since their contribution is rather small and slowly varies with energy. That is why it is convenient to derive exactly a few significant poles and the corresponding input and output vectors by applying the described pole-search algorithm until it converges to each pole. In other words, to determine $N$ resonant poles one has to utilize the algorithm at least $N$ times so that each time a new resonance is determined with sufficient precision. The impact of the other distant poles can be taken into account by the means of the background scattering matrix $\mathbb{S}_b$ that does not possess any sharp resonant behavior.

Note that the eigenvectors in X are generally not
orthogonal, and $\langle O_n|$ is not defined as the Hermitian conjugate of vector $|O\rangle$, instead, it is a row vector of $\mathrm{X}^{-1}$. In the same manner, $|I_n\rangle$ is defined as a column vector in $\mathrm{Y}^{-1}$.
Both $\mathrm{X}$ and $\mathrm{Y}$ are dependent on the energy $\omega'$, but if $\omega'$ is located in the vicinity of some resonant energy $\omega_n$ the corresponding product of the output and input resonant vectors $\left|O_{n}\right\rangle \left\langle I_{n}\right|$ is a calculated pole residual.

Here it is important to underline, that the described algorithm could be implemented for any other complex parameter space. For example, one can instead find poles of the scattering matrix in the complex space of the wavevector's $z$-projection $k_z$.}

 \bibliographystyle{elsarticle-num} 
 \bibliography{JAB_library}

\end{document}